\documentclass[aps,prl,twocolumn,showpacs,superscriptaddress]{revtex4-1}  

\usepackage{graphicx}  
\usepackage{dcolumn}   
\usepackage{bm}        
\usepackage{amssymb}   
\usepackage{mathtools}
\usepackage{etoolbox}

\appto\normalsize{\belowdisplayshortskip=\belowdisplayskip}
\appto\small{\belowdisplayshortskip=\belowdisplayskip}
\appto\footnotesize{\belowdisplayshortskip=\belowdisplayskip}
\usepackage[utf8]{inputenc}
\DeclareMathSizes{5}{13}{7}{3}

\begin{document}

\title{Topological magnetic phase transition in Eu-based A-type antiferromagnets}
\author{Eliot Heinrich}
\address{Department of Physics, Boston College, 140 Commonwealth Avenue, Chestnut Hill, Massachusetts 02467, USA}
\author{Thore Posske}
\address{
 I. Institute for Theoretical Physics, Universit{\"a}t Hamburg, Notkestraße 9, 22607 Hamburg, Germany}
\address{The Hamburg Centre for Ultrafast Imaging, Luruper Chaussee 149, 22761 Hamburg, Germany}
\author{Benedetta Flebus}
\address{Department of Physics, Boston College, 140 Commonwealth Avenue, Chestnut Hill, Massachusetts 02467, USA}

\date{\today}

\begin{abstract}

Recently, a colossal magnetoresistance (CMR) was observed in EuCd$_2$P$_2$ -- a compound that does not fit the conventional mixed-valence paradigm. Instead, experimental evidence points at a resistance driven by strong magnetic fluctuations within the two-dimensional ($2d$) ferromagnetic (FM) planes of the layered antiferromagnetic (AFM) structure. While the experimental results have not yet been fully understood, a recent theory relates the CMR to a topological vortex-antivortex unbinding, i.e., Berezinskii-Kosterlitz-Thouless (BKT), phase transition. Motivated by these observations, in this work we explore the magnetic phases hosted by a microscopic classical magnetic model for EuCd$_2$P$_2$, which easily generalizes to other Eu A-type antiferromagnetic compounds. Using Monte Carlo techniques to probe the specific heat and the helicity modulus, we  show that our model can exhibit a vortex-antivortex unbinding phase transition. We find that this phase transition displays the same sensitivity to in-plane magnetization, interlayer coupling, and easy-plane anisotropy that is observed experimentally in the CMR signal, providing qualitative numerical evidence that the effect is related to a magnetic BKT transition.

\end{abstract}

\maketitle

\section{Introduction}

For decades, the study of magnetic phase transitions has focused on a handful of paradigmatic models. One such model is the $2d$ XY model, also called the planar rotor or $O(2)$ model. This system has a continuous $2d$ rotational symmetry, and thus supports long-wavelength fluctuations that prevent spontaneous symmetry-breaking by the Mermin-Wagner theorem \cite{mermin1966}. Nonetheless, in 1973,  Berezinskii, Kosterlitz, and Thouless (BKT) \cite{kosterlitz1973, kosterlitz1974, berezinskii11971} showed that the XY model undergoes a magnetic phase transition outside of the Landau symmetry-breaking paradigm, instead driven by the unbinding of (anti)vortex pairs and subsequent proliferation of these topological defects. This seminal work prompted intense numerical \cite{nelson1977, loft1987, hasenbusch2005, komura2012} and experimental \cite{bishop1978, pargellis1994, hadzibabic2006, hu2020, resnick1981, hu2020} investigations.

The recent discovery of van der Waals (vdW) magnets has provided a truly $2d$ platform to test magnetic phenomena in lower dimensions \cite{burch2018}, which has prompted significant work to find magnetic realizations of the BKT transition \cite{bedoyapinto2021, johansen2019, kim2019, wiedenmann1981, seifert2022}. Even multilayer systems can display properties characteristic of BKT transitions, provided the interlayer coupling is small \cite{hikami1980, hirakawa1982, willa2019, cornelius1986, bramwell1995, janke1990}. One such proposed magnetic platform is EuCd$_2$P$_2$, an $S=7/2$ A-type antiferromagnet on a trigonal lattice with easy-plane anisotropy \cite{wang2021}, shown in Fig. \ref{fig:lattice}. A recent experimental work reported a temperature-dependent colossal spike in the magnetoresistance of this compound  that does not fit into the typical mixed valence paradigm of manganese perovskite materials \cite{wang2021}. The spike appears above the Neél temperature  where, in spite of the weak interlayer coupling, easy-plane physics can be accessed. The spike disappears when the interlayer coupling is increased e.g., via chemical tuning of nonmagnetic ligands, suggesting that the underlying mechanism stems from $2d$ phenomena. Additionally, the spike is strongly suppressed by the presence of an external magnetic field. The suppression of the magnetoresistance is paralleled by the suppression of the dynamic magnetic susceptibility at the same temperature, indicating that the mechanism behind the colossal magnetoresistance is related to strong magnetic fluctuations. An intriguing explanation for the observed magnetoresistive spike invokes electron scattering on magnetic defects at a topological transition temperature, where magnetic defects proliferate exponentially \cite{flebus2021}. As a magnetic field \cite{jose1977} and $3d$ ordering \cite{hu2020, janke1990} are known to be detrimental to the BKT transition, this theory appears to be consistent with the known behavior of EuCd$_2$P$_2$. 

In this work, we  build up a magnetic model of EuCd$_2$P$_2$, which can be easily generalized to related compounds, and explore its thermodynamic properties with Monte Carlo techniques.  We find that the system displays three phases: i) a high-temperature paramagnetic phase, ii) a mid-temperature quasi-ordered phase characterized by bound vortex-antivortex pairs (as in Fig. \ref{fig:lattice}(b)), and iii) a symmetry-breaking ferromagnetic low-temperature phase. We show that the higher-temperature phase transition between the first two phases, i.e., i) and ii), is of the topological BKT type, whereas the lower-temperature one is a symmetry-breaking transition captured by Landau theory. We also find that the BKT transition is suppressed by in-plane magnetic fields and interlayer coupling, which is qualitatively consistent with the behavior of the experimental magnetoresistive spike, thereby providing numerical qualitative evidence of a connection between the observed CMR and a magnetic BKT transition. 

However, we can not quantitatively reproduce the anisotropic magnetic susceptibility of $\text{EuCd}_{2}\text{P}_{2}$\cite{wang2021}, possibly due to magnon-phonon interactions or lattice strain effects that are not accounted for in our model. Thus, we cannot provide a quantitative estimate of the BKT transition temperature, nor of the dependence of the magnetoresistance on experimental parameters, without further experimental investigations.

This work is organized as follows: In Sec. II, we introduce our model and the thermodynamic observables used to characterize the phase transitions. In Sec. III, we discuss our numerical results, and in Sec. IV, we present our conclusions and an outlook. A discussion of the allowed single-ion anisotropic terms based on the lattice symmetry is presented in Appendix A, and the details of the Monte Carlo simulations are discussed in detail in Appendix B.

\section{Model and Methods} 

\begin{figure}[t!]
    \includegraphics[width=\linewidth]{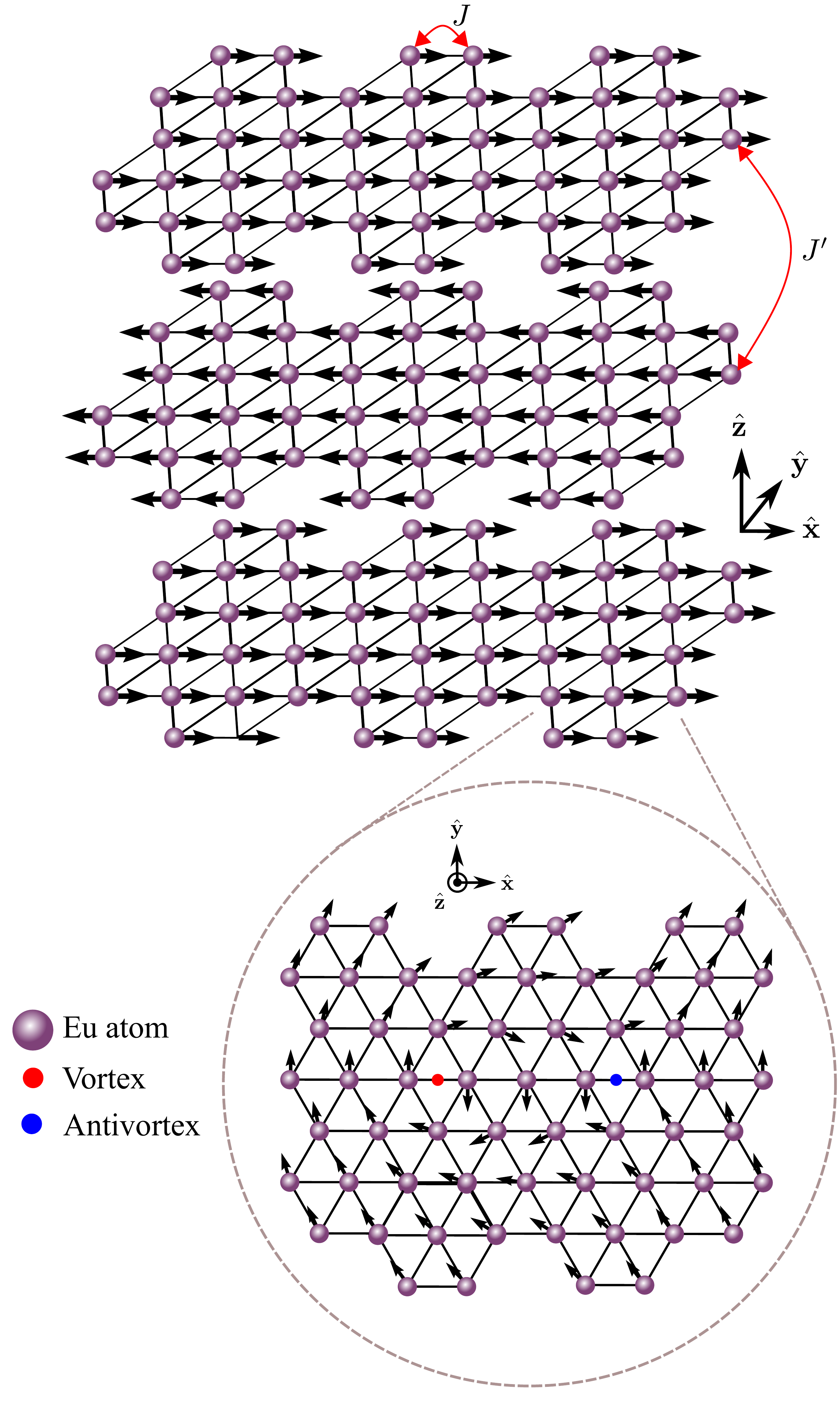}
    \caption{Magnetic crystal structure of Eu-based A-type antiferromagnets.   $\text{Eu}^{2+}$ spins are arranged in a trigonal layer in the $xy$-plane. The spins are aligned ferromagnetically in the $xy$-plane whereas the spins in adjacent $\text{Eu}$ planes along the $z$-axis are aligned antiferromagnetically. The intralayer (interlayer) coupling $J$ ($J'$). Inset:  Within each layer, the easy-plane anisotropy stabilizes (anti)vortices as topological excitations with cores at the red (blue) markers. Below a transition temperature, these defects form in vortex-antivortex pairs, and above the transition temperature, the pairs unbind, giving way to the Berezinskii-Konsterlitz-Thouless transition. }
    \label{fig:lattice}
\end{figure}

Eu A-type antiferromagnetic compounds are layered structures whose magnetic moments are carried by the Eu$^{2+}$ atoms. The spins within each magnetic layer are ferromagnetically aligned, while neighboring planes are coupling antiferromagnetically via nonmagnetic atoms, as shown in Fig.~\ref{fig:lattice}(a).  Experimental data suggest an easy-plane and a sixfold magnetocrystalline anisotropies \cite{wang2021}. A minimal Hamiltonian model accounting for  exchange and Zeeman interactions and magnetocrystalline anisotropy can be then written as

\begin{align}\label{fullhamiltonian}
    \mathcal{H} =& -J\sum\limits_{\langle i, j \rangle, \ell} \mathbf{S}_{i,\ell} \cdot \mathbf{S}_{j,\ell} + J' \sum\limits_{i,\langle \ell, \ell' \rangle} \mathbf{S}_{i,\ell} \cdot \mathbf{S}_{i,\ell'} \nonumber \\
    &+ K_2 \sum\limits_{i,\ell} (S_{i,\ell}^z)^2 + K_6\sum\limits_{i,\ell} \sin^6(\theta_{i,\ell})\cos(6\phi_{i,\ell}) \nonumber \\
    &- \mathbf{B} \cdot \sum\limits_{i,\ell} \mathbf{S}_{i,\ell}
\end{align}
where $\mathbf{S}_{i,\ell} = (\cos(\phi_{i,\ell})\sin(\theta_{i,\ell}), \sin(\phi_{i,\ell})\sin(\theta_{i,\ell}), \cos(\theta_{i,\ell}))$ is a normalized vector with $\theta_{i,\ell}$ and $\phi_{i,\ell}$ the polar and azimuthal angles, respectively, relative to the Cartesian coordinate system sketched in Fig.~\ref{fig:lattice}. The vectors $\mathbf{S}_{i,\ell}$ model the spin magnetic moment of $\text{Eu}^{2+}$ ions on a trigonal lattice. While $\text{Eu}^{2+}$ ions  have spin angular momentum  $S=7/2$, here we set $|\mathbf{S}| = 1$ and absorb the factor $S$ into the coupling constants. The indices $i$ and $j$ label the $2d$ intralayer position and $\ell$ and $\ell'$ label the layer. The sum taken over $\langle i,j \rangle (\langle \ell, \ell' \rangle)$ indicates a sum over nearest-neighbor intrayer (interlayer) pairs. Here $\mathbf{B}$ is the applied magnetic field, while  $J>0$ and $J'>0$ parameterize, respectively,  the strength of the intralayer FM and interlayer AFM Heisenberg exchange coupling.  $K_{2}>0$ and $K_{6}>0$ are the constants governing the strength of the local easy-plane and sixfold anisotropy, respectively. While our analysis focuses on the aforementioned anisotropies, we report a systematic analysis of magnetocrystalline anisotropies allowed by symmetry in Appendix A.

Depending on the relative strength of the magnetic interactions, the model~(\ref{fullhamiltonian}) can host several magnetic phases, as we will show in detail. To characterize these phases, we primarily employ two observables: i) the specific heat, and ii) the helicity modulus. The specific heat can be calculated as

\begin{align}\label{specificheat}
	c(T) = \frac{1}{N} \frac{\langle \mathcal{H}^2 \rangle - \langle \mathcal{H} \rangle^2}{k_B T^2},
\end{align}
where $N = L^2$ is the number of magnetic moments with $L$ the linear size of the system, $\langle \cdot \rangle$ represents a thermal average with respect to the Boltzmann distribution, $T$ is the temperature and $k_{B}$ is the Boltzmann factor.

Discontinuities in the specific heat $c(T)$ or its derivatives are a signature of phase transitions or crossover phenomena. However, they do not carry information about the specifics of the transition nor crossover. To properly identify a phase transition as a BKT transition, we employ a standard analysis of the helicity modulus $\Upsilon$ following Refs. \cite{hasenbusch2005, hsieh2013}. The helicity modulus is a thermodynamic quantity that accounts for the system's response to an infinitesimal phase twist along a certain direction. To introduce it formally, we define the following transformation: 

\begin{align}\label{phasetwist}
    \sum\limits_{\langle i, j \rangle} \mathbf{S}_i \cdot \mathbf{S}_j \rightarrow \sum\limits_{\langle i, j \rangle} \mathbf{S}_i \cdot R_z(\Delta \hat{\boldsymbol \alpha}_{ij} \cdot \mathbf{\hat{x}}) \mathbf{S}_j
\end{align}
where $\hat{\boldsymbol \alpha}_{ij}$ is the unit vector pointing from site $i$ to $j$ and $R_z(\phi)$ is a rotation matrix representing a clockwise rotation about the $\mathbf{\hat{z}}$-axis by angle $\phi$. The Hamiltonian correspondingly transforms $\mathcal{H} \rightarrow \mathcal{H}(\Delta)$. The helicity modulus, in turn, is the second derivative of the thermodynamic free energy in response to this transformation:

\begin{align}\label{helicitymodulus}
    \Upsilon(T) =& \frac{1}{N}\frac{\partial^2 F}{\partial \Delta^2}\bigg\vert_{\Delta = 0} \nonumber \\
    =& \frac{1}{N} \left\langle \frac{\partial^2 \mathcal{H}(\Delta)}{\partial \Delta^2}\bigg\vert_{\Delta = 0}\right\rangle \nonumber \\
    &- \frac{1}{N \beta} \left[ \left\langle \frac{\partial \mathcal{H}(\Delta)}{\partial \Delta} \bigg\vert_{\Delta = 0} \right\rangle^2 - \left\langle \left( \frac{\partial \mathcal{H}(\Delta)}{\partial \Delta}\right)\bigg\vert_{\Delta = 0}^2 \right\rangle \right].
\end{align}

In the thermodynamic limit, the helicity modulus obeys the Nelson-Kosterlitz universal relation \cite{nelson1977}:

\begin{align}
    \lim\limits_{T \rightarrow T_{BKT}^-} \Upsilon(T) &= \frac{2 T_{BKT}}{\pi},\label{nelsonkosterlitz} \\
    \lim\limits_{T \rightarrow T_{BKT}^+} \Upsilon(T) &= 0. \label{nelsonkosterlitz1}
\end{align}
where the limits are taken from above and below in Eqs.~(\ref{nelsonkosterlitz}) and~(\ref{nelsonkosterlitz1}), respectively.
One can extract the critical temperature $T_{BKT}$ and  the BKT character of a phase transition by identifying the temperature at which the helicity modulus jumps discontinuously from $2T_{BKT}/\pi$ to $0$. In practice, finite size effects must be taken into account for this to be applicable. The correlation length above $T_{BKT}$ of the $2d$ XY model is characteristically exponential: $\xi \sim e^{a/(T_{BKT})^{1/2}}$ \cite{maccari2018}. Since finite size effects become important when $\xi \sim L$, we can see that the finite size correction to $T_{BKT}$ decays logarithmically with $L$. We therefore conduct a finite-size scaling analysis following Ref. \cite{schultka1994}. The helicity modulus in the thermodynamic limit $L \rightarrow \infty$ is related to the finite-size helicity modulus $\Upsilon(T, L)$ by \cite{hsieh2013, maccari2018, hasenbusch2005}

\begin{align}\label{finitesizeupsilon}
    \Upsilon(L \rightarrow \infty, T_{BKT}) = \frac{\Upsilon(L, T_{BKT})}{1 + \left[2 \log (L/L_0)\right]^{-1}} + O\left(\frac{1}{\log L^3} \right),
\end{align}
where $L_0$ is a free system-dependent parameter. From Eq. (\ref{finitesizeupsilon}) and Eq.  (\ref{nelsonkosterlitz}), one finds that the family of curves $\Upsilon(L, T)/(1+[2\log(L/L_0)]^{-1})$ intersects at $T_{BKT}$ with the critical line with slope $2/\pi$. By tuning $L_0$ and searching for such an intersection, we can determine the critical temperature and BKT character of the phase transition. This procedure is carried out as follows: for each $L$ sampled, we define $T_{BKT}(L)$ by the temperature at which $\Upsilon(L,T_{BKT}(L))/\left[1 + (2\log(L/L_0))^{-1}\right] = 2T_{BKT}(L)/\pi$. Then, the quantity $\sigma_{T_{BKT}} = \langle T_{BKT}(L)^2 \rangle - \langle T_{BKT}(L) \rangle^2$, where $\langle \cdot \rangle$ is the average over every $L$, is minimized with respect to $L_0$ to determine the $L_0$ that gives the best intersection of each $\Upsilon(L,T)/\left[1 + (2\log(L/L_0))^{-1}\right]$ on the critical line.

\begin{figure*}[t]
    \includegraphics[width=\textwidth]{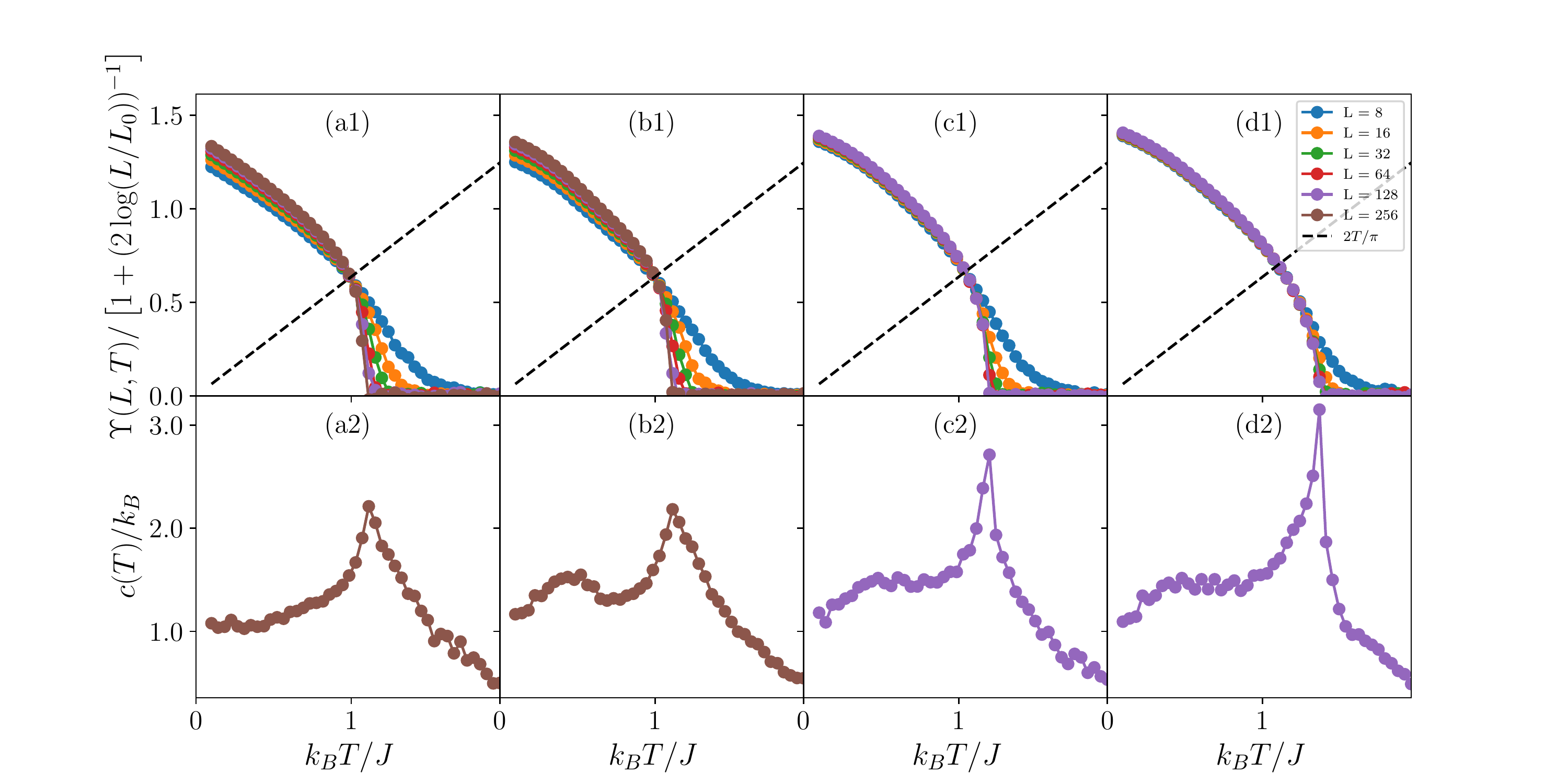}
    \caption{We explore the presence or absence of a BKT transition as function of the parameters of our model~\eqref{fullhamiltonian}  by plotting the normalized helicity modulus~(\ref{finitesizeupsilon}) (upper panels) and the specific heat~(\ref{specificheat}) (lower panels) for: (a1,a2) A monolayer with finite easy-plane anisotropy, i.e., $K_2/J = 0.25$ while setting $K_6 = J' = |\mathbf{B}| = 0$; (b1, b2) A monolayer with finite easy-plane and sixfold anisotropies, i.e., $K_2/J = 0.25$, $K_6/J = 0.2$, while setting  $J' = |\mathbf{B}| = 0$; (c1,c2) A tetralayer with  weak  interlayer coupling and finite easy-plane and sixfold anisotropies, i.e., $K_2/J = 0.25$, $K_6/J = 0.2$, $J'/J = 0.01$, while setting $|\mathbf{B}| = 0$; (d1,d2) A multilayer with four layers with stronger interlayer coupling and finite easy-plane and sixfold anisotropies, i.e., $K_2/J = 0.25$, $K_6/J = 0.2$, $J'/J = 0.1$, while setting $|\mathbf{B}| = 0$. 
    A second peak in the specific heat at $k_B T / J \approx 0.5$ in (b2) signals a second phase transition. In (c1),   the normalized helicity modulus curves no longer satisfactorily intersect on the Nelson-Kosterlitz line, signaling the suppression of the BKT transition. In (a1), (b1), (c1), and (d1) the black dashed line corresponds to the critical line on which the normalized $\Upsilon(L,T)$ curves must intersect to satisfy the Nelson-Kosterlitz relation~(\ref{nelsonkosterlitz}). The critical temperature $T_{BKT}$ is the temperature at which the curves intersect on the critical line. }
    \label{fig:overview}
\end{figure*}

\section{Results}

\begin{figure*}
    \setlength{\lineskip}{-3.5em}
    \centering
    \includegraphics[width=\linewidth]{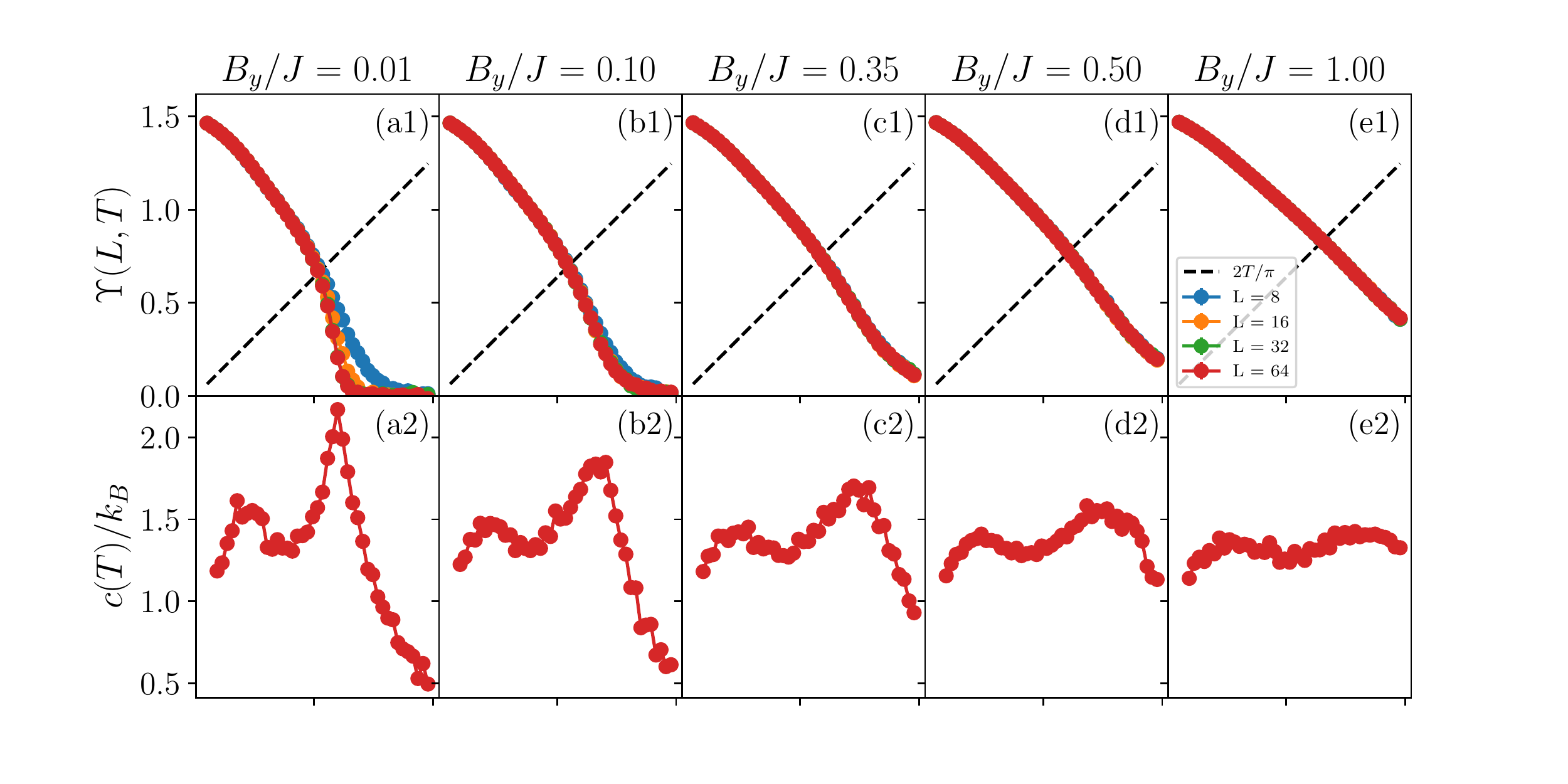}
    \includegraphics[width=\linewidth]{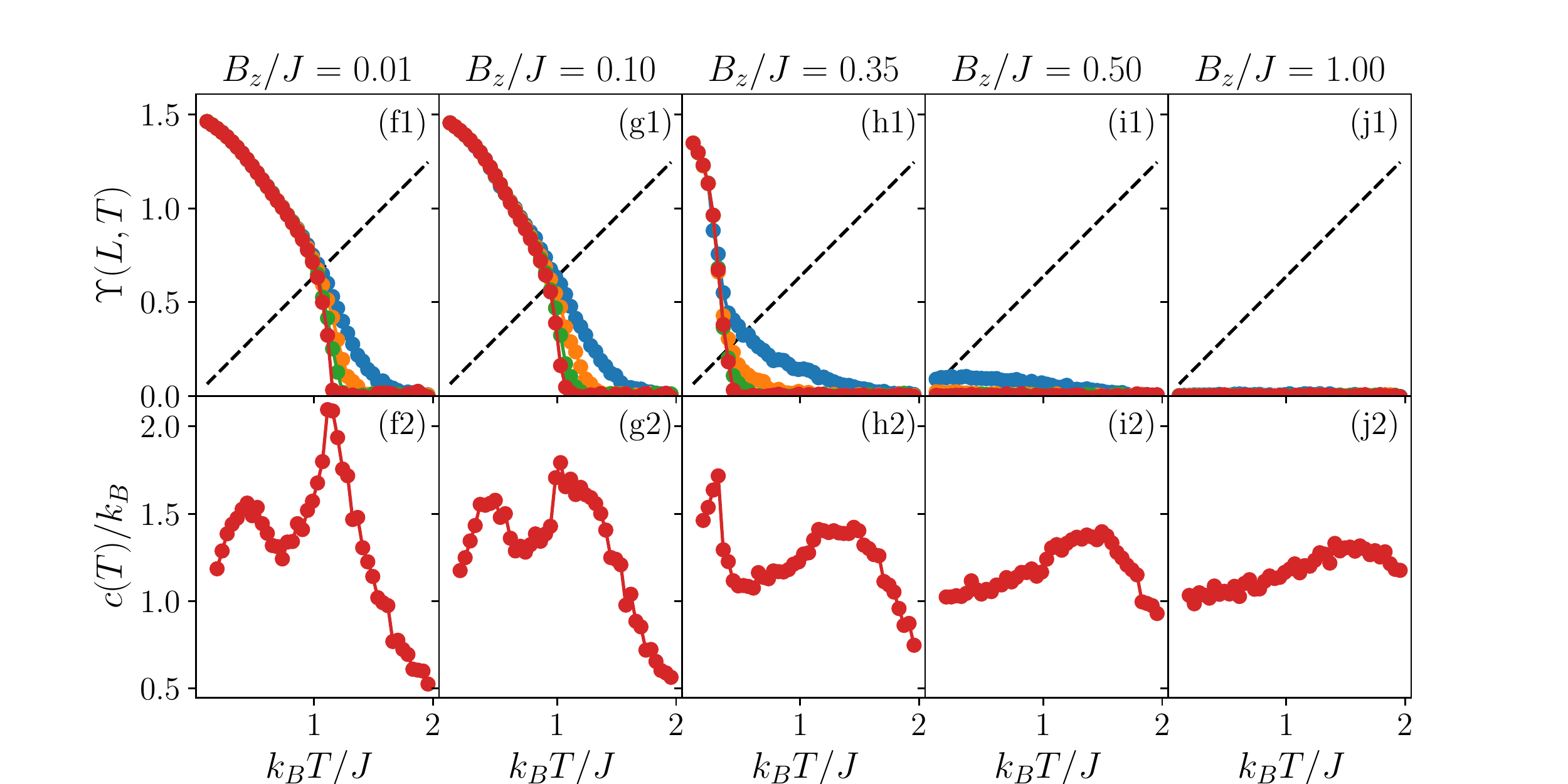}
    \caption{To explore the dependence of the BKT transition on the strength and direction of the external magnetic field, we plot the un-normalized $\Upsilon(L,T)$ and specific heat $c(T)/k_B$ against magnetic field strengths in-plane and out-of-plane for the (a-e) and (f-j) rows, respectively. The in-plane magnetic field quickly suppresses the BKT transition, as can be seen from the higher temperature peak becoming smaller between (a2) and (d2) and the failure to produce a non-trivial intersection on the Nelson-Kosterlitz line in (a1) though (e1). The out-of-plane magnetic field also suppresses the BKT transition, diminishing the high-temperature peak in the specific heat, but only once the magnetic field turns the average magnetization out of plane. In this case, the low-temperature symmetry-breaking transition is suppressed as well,  possibly due to  $\sin^6(\theta)\rightarrow 0$ in Eq. (\ref{fullhamiltonian}). }
    \label{fig:figB}
\end{figure*}

In this section, we consider several limits of our model Eq.~(\ref{fullhamiltonian}) and investigate its phase transitions via annealed spin-cluster Monte Carlo simulations. Imposing periodic boundary conditions in-plane and out-of-plane, for $L = 8, 16, 32, 64, 128$, we perform $1.5\times10^{6}$ Wolff cluster updates \cite{wolff1989} to equilibrate, followed by $5\times10^{4}$ further cluster updates, sampling the helicity modulus and specific heat after each update. For $L = 256$, we perform $5 \times 10^{5}$ updates to equilibrate and take $5 \times 10^{4}$ samples. More details on the simulation are discussed in Appendix B.  

We start by considering a $2d$ classical Heisenberg ferromagnet, which our model (\ref{fullhamiltonian}) reduces to for vanishing onsite anisotropy, magnetic field, and interlayer coupling, i.e.,  $K_2 = K_6 = J' = |\mathbf{B}| = 0$. This model has a continuous rotational symmetry, which, according to the Mermin-Wagner Theorem \cite{mermin1966}, prevents the spontaneous breaking of this symmetry by the formation of infrared divergent Goldstone modes, i.e., spin waves. While this system has no symmetry-breaking phase transition, the presence or absence of a phase transition in the $2d$ classical Heisenberg ferromagnet has remained an open and controversial question \cite{blte2002, schmoll2021, stanley1966}. Monte Carlo studies, as in the present work, are obligated to introduce a finite size to the system, which in turn places an upper cutoff for the spin-wave wavelength. Therefore, a symmetry-breaking phase transition may be observed as a numerical artifact, even if such a transition is formally forbidden by the Mermin-Wagner theorem. 

Let us now allow for a finite in-plane anisotropy, i.e., $K_2 \neq 0$. The easy-plane anisotropy reduces the symmetry of the system from $O(3)$ to $O(2)$ by favoring spin configurations in the $xy$-plane when $K_2 > 0$ . The model still has a continuous symmetry, and thus no symmetry-breaking phase transition in the thermodynamic limit. However, the easy-plane anisotropy stabilizes meron topological defects \cite{lu2020}. When the meron radius becomes smaller than a lattice spacing, merons manifest as vortices, characteristic of the XY model \cite{gouva1989}, which is understood to exhibit a topological phase transition. The proliferation of topological defects drives a phase transition from a low-temperature quasi-ordered phase to a high-temperature paramagnetic phase according to the theory of Berezinskii, Kosterlitz, and Thouless (BKT) \cite{kosterlitz1973, kosterlitz1974, berezinskii11971}. Taking $K_6 = J' = |\mathbf{B}| = 0$ and $K_2/J = 0.25$, this transition is indicated by the cusp in the specific heat, as can be seen in Fig.~\ref{fig:overview}(a2). For $L_0 = 0.43$, the normalized helicity modulus satisfies the Nelson-Kosterlitz universal relation at $k_B T/J = 0.994$, as can be seen in Fig.~\ref{fig:overview}(a1), indicating that the transition is of the BKT type. 

The observed anisotropic magnetic response suggests a magnetocrystalline field which breaks the continuous rotational symmetry \cite{wang2021}. We use the form proportional to $K_6$ in Eq.~\ref{fullhamiltonian}, though the $D_{3d}$ lattice point symmetry of EuCd$_2$P$_2$ allows for other onsite anisotropic fields. In Appendix A, we will explore other allowed magnetocrystalline anisotropies, while here we focus on the sixfold term proposed. The introduction of the sixfold term proportional to $K_6$ breaks the continuous rotational symmetry of Eq.~(\ref{fullhamiltonian}), reducing the symmetry group to $\mathbb{Z}_2 \times \mathbb{Z}_6$. Critically, this symmetry group is discrete: thus, the system can and indeed does support a low-temperature symmetry-breaking phase that gives way to an intermediate bound-vortex BKT phase at $k_B T_1/J \approx 0.5$ and finally to the paramagnetic high-temperature phase at $T_{BKT}$ \cite{jose1977, kumano2013, miyajima2021}, as shown in Fig.~\ref{fig:overview}(b). The addition of a finite magnetocrystalline field $K_6/J = 0.2$ does not appear to change the BKT character or appreciably change the BKT transition temperature in Fig. \ref{fig:overview}(b1), but it does introduce a second peak in the specific heat at $k_B T/J \approx 0.5$ in Fig. \ref{fig:overview}(b2).

\begin{figure}[t]
    \centering
    \includegraphics[width=0.5\textwidth]{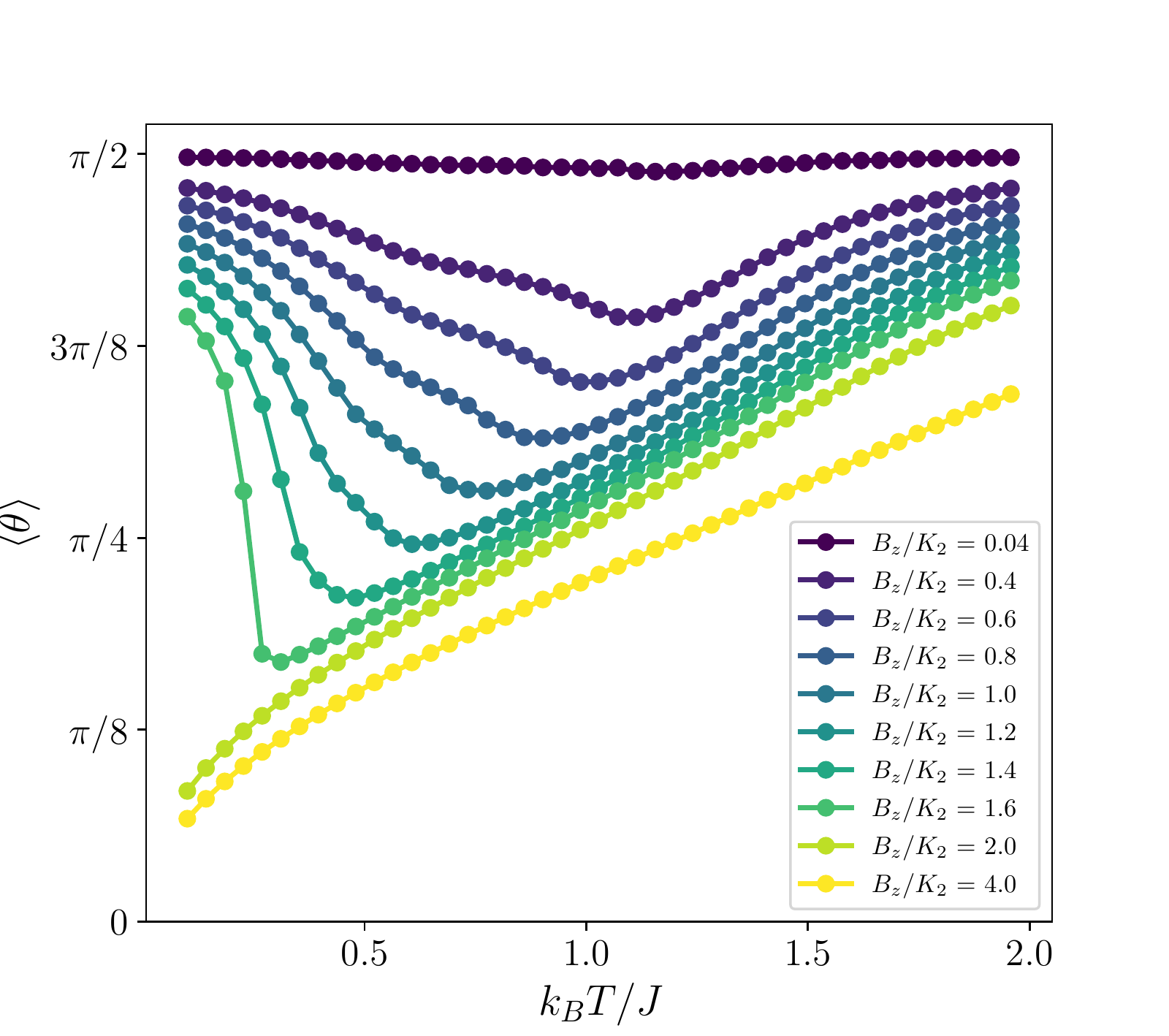}
    \caption{The average polar angle $\langle \theta \rangle$ as a function of temperature for representative out-of-plane magnetic field strengths. When the out-of-plane magnetic field is larger than the easy-plane anisotropy, the low-temperature magnetization abruptly jumps out-of-plane. Comparing with Fig. \ref{fig:figB}, we observe that this jump is accompanied by the disappearance of the symmetry-breaking phase transition. }
    \label{fig:polarangle}
\end{figure}

Allowing for finite interlayer coupling, i.e., $J' \neq 0$, introduces a $3d$ magnetic order. For weak $J'/J = 0.01$, both the helicity modulus in Fig.~\ref{fig:overview}(c1) and cusp in the specific heat in Fig.~\ref{fig:overview}(c2) suggest that the BKT transition still takes place. As the interlayer coupling becomes stronger, e.g., $J'/J = 0.1$, the BKT character of the high-temperature phase transition vanishes, as indicated by the lack of a suitable $L_0$ that non-trivially satisfies the Nelson-Kosterlitz relation in Fig.~\ref{fig:overview}(d1). Still, a magnetic phase transition is signaled by the behavior of the specific heat, shown in Fig.~\ref{fig:overview}(d2). In $3d$, a spontaneous magnetization is no longer forbidden by the Mermin-Wagner theorem, so the system has an ordered to disordered phase transition which can be described in the paradigm of Landau theory. This order to disorder transition seems insensitive to the sign of $J'$; setting $J' < 0$, which corresponds to a ferromagnetic interlayer coupling, produces the same features in the specific heat. This can be  understood as, for an even number of layers, the state with ferromagnetic interlayer order is related to the state with antiferromagnetic interlayer coupling by a reflection of $\mathbf{S}_{i,\ell} \rightarrow -\mathbf{S}_{i,\ell}$ for even $\ell$. This crossover from $3d$ to $2d$ appears to happen continuously in $J'$, suggesting BKT behavior may be present for sufficiently small $J'$ \cite{janke1990}.

Finally, we explore the dependence of the magnetic phase transition on the external magnetic field, by considering both  in-plane, i.e., $\mathbf{B} = B_y \hat{\mathbf{y}}$, and  out-of-plane, $\mathbf{B} = B_z \hat{\mathbf{z}}$, magnetic fields. For an in-plane magnetic field, we see that -- even for relatively small values of $B_{y}/J$ -- the preferred direction of alignment for the spins destabilizes vortex excitations and suppresses the BKT character of the phase transition. This is numerically evidenced by the lack of an $L_0$ which satisfies the Nelson-Kosterlitz universal relation, as seen in Fig.~\ref{fig:figB}(a1) through \ref{fig:figB}(e1). As the in-plane magnetic field becomes stronger, both phase transitions appear to disappear, as can be seen in Fig.~\ref{fig:figB}(a2) through Fig.~\ref{fig:figB}(e2).  

 On the other hand, weak out-of-plane magnetic fields do not appear to suppress the BKT transition, as shown in Fig.~\ref{fig:figB}(f1) and \ref{fig:figB}(g1). However, once $B_z/K_2 > 1.6$, the ground state magnetization turns out of plane, as can be seen in Fig.~\ref{fig:polarangle}, and the signature of the symmetry-breaking transition disappears in the specific heat. This is signalled by the lower temperature anomaly disappearing in the specific heat, which is visible in Fig.~\ref{fig:figB}(f2) through \ref{fig:figB}(h2) disappearing in Fig.~\ref{fig:figB}(i2) and \ref{fig:figB}(j2). We explain this in terms of the average polar angle $\langle \theta \rangle = \arccos\sum_{i,\ell} \langle S_{i,\ell}^z \rangle$; as $\langle \theta \rangle \rightarrow 0$, the factor of $\sin^6(\theta_i)$ in the sixfold anisotropic term in Eq. (\ref{fullhamiltonian}) goes to zero, setting the spin wave gap to zero, and long-range fluctuations become allowed again. This causes the spontaneous symmetry-breaking transition to disappear in the specific heat. The BKT character is also lost when the magnetization jumps out of plane, as can be seen from Fig.~\ref{fig:figB}(h1) through \ref{fig:figB}(j1).

\section{Conclusion and outlook} 

In this work, we have built a microscopic magnetic model for a layered Europium based A-type antiferromagnet, paying special attention to EuCd$_2$P$_2$, and explored the resulting magnetic phase transitions as function of  easy-plane and sixfold-symmetric magnetocrystalline anisotropies and  interlayer coupling. 

While EuCd$_2$P$_2$ displays a colossal magnetoresistance an order of magnitude higher than unoptimized manganates \cite{wang2021}, EuCd$_2$As$_2$ and EuCd$_2$Sb$_2$ are also known to exhibit a spike in magnetoresistance at a critical temperature, though the effect is much less pronounced than in EuCd$_2$P$_2$. We note that a possible factor distinguishing EuCd$_2$P$_2$ from the other listed materials is the magnitude of the magnetic anisotropy: e.g., EuCd$_2$P$_2$ is known to have a higher easy-plane anisotropy than EuCd$_2$As$_2$ and weaker interlayer coupling, and thus $2d$ easy-plane physics may be enhanced. We indeed find that the easy-plane anisotropy stabilizes topological defects that can mediate a $2d$ BKT-like transition in our model. A sixfold magnetocrystalline field circumvents the restrictions of the Mermin-Wagner theorem and introduces a second phase transition associated with finite-temperature spontaneous magnetization. The topological phase transition appears to be stable to out-of-plane magnetic fields weaker than the easy-plane anisotropy. At higher out-of-plane magnetic field strengths, the average polar angle $\langle \theta \rangle$ goes to zero and the sixfold term no longer contributes, thereby suppressing the symmetry-breaking magnetic phase transition. On the other hand, an in-plane magnetic field suppresses the topological magnetic phase transition, even for relatively small values of $B_y/J$. Our results could be corroborated by systematic measurements of the CMR dependence on the strength and direction of the field.
Furthermore, we find  that, while the BKT transition is sensitive to the interlayer coupling, it does not  vanish upon introduction of small $J'$.

Our results are consistent with the hypothesis that the observed CMR is associated with the emergence of a BKT transition and the consequent proliferation of magnetic defects \cite{flebus2021}. In particular, the qualitative influence of the easy-plane anisotropy, interlayer coupling, as well as symmetry-breaking phase transition as a consequence of the sixfold magnetocrystalline anisotropy, is captured correctly. Yet, our model fails to reproduce the measured anisotropic magnetic susceptibility at low temperatures \cite{wang2021}; the observed magnetization increases more quickly in the $x$ direction relative to the $y$ directions with increasing temperature, whereas our model produces a similar response in the low-temperature regime regardless of the value of $K_6$ associated with the sixfold anisotropic term. Further, the inability to qualitatively fit to experiment leaves us unable to estimate the coupling constants which appear in our model, and thus a numerical comparison to experiment is not yet possible. This discrepancy suggests that further experimental studies on the magnetic behavior of EuCd$_2$P$_2$ are required. 

Furthermore, the dependence of the coupling between electronic and magnetic degrees of freedom on the magnetic phases should be addressed. We leave to future work the task of capturing the complexity of the bad-metal behavior of EuCd$_2$P$_2$ within a representative electronic model \cite{wang2021}. 

\section{Acknowledgments}

The authors thank F. Tafti for insightful discussions and acknowledge computational support from the Andromeda computing cluster at Boston College. E.H. acknowledges the Universit{\"a}t Hamburg’s Next Generation Partnership funded under the
Excellence Strategy of the Federal Government and the
L{\"a}nder. T.P. acknowledges funding by the DFG (project no. 420120155). B.F. acknowledges support from the National Science Foundation under Grant No. NSF DMR- 2144086.

\appendix

\section*{Appendix A: Single-ion anisotropies}

In this work, we focused on a single-ion anisotropy of the form $K_2 \cos(\theta)^2 + K_6 \sin^6(\theta)\cos(6\phi)$, where $\theta$ and $\phi$ are, respectively, the polar and azimuthal angles of a spin. However, the lattice of EuCd$_2$P$_2$ has $D_{3d}$ point symmetry. Thus,  any magnetocrystalline anisotropy that is invariant under the action of $D_{3d}$ is in principle allowed. The $D_{3d}$ point symmetry group is generated by a rotation about the $\hat{\mathbf{z}}$-axis by $2\pi/3$, a reflection across the $\hat{\mathbf{x}}$-axis, and a rotation about the $\hat{\mathbf{y}}$-axis by $\pi$. Seeking polynomials in $\{S_x, S_y, S_z\}$ that are invariant under these generators, we find, up to $O(S^6)$, the terms in Table. I.

\begin{table}[h!]
\begin{center}
    \begin{tabular} {c c c c c}
        & Cartesian & Polar \\
        $O(S^2)$: & $S_z^2$ & $\cos^2(\theta)$ & (1) \\
                  & $S_x^2 + S_y^2$ & $\sin^2(\theta)$ & (2) \\
        $O(S^4)$: & $S_z^4$ & $\cos^4(\theta)$ & (3) \\
                  & $S_x^4 + S_y^4 + 2S_x^2 S_y^2$ & $\sin^4(\theta)$ & (4) \\
        $O(S^6)$: & $S_z^6$ & $\cos^6(\theta)$ & (5) \\ 
                  & $S_x^6 - 15 S_x^4 S_y^2 + 15 S_x^2 S_y^4 - S_y^6$ & $\sin^6(\theta)\cos(6\phi)$ & (6)\\
                  & $S_x^2 S_z^4 + S_y^2 S_z^4$ & $\sin^2(\theta)\cos^4(\theta)$ & (7)\\
                  & $S_x^5 S_z - 2 S_x^3 S_y^2 S_z - 3 S_x S_y^4 S_z$ & $\sin^5(\theta) \cos(\theta) \cos(3\phi)$ & (8)\\
                  & $S_x^3 S_z^3 - 3 S_x S_y^2 S_z^3$ & $\sin^3(\theta)\cos^3(\theta)\cos(3\phi)$ & (9) \\
    \label{table:onsiteterms}
    \end{tabular}
    \caption{The single-ion anisotropy terms allowed by $D_{3d}$ point symmetry in spin cartesian and polar coordinates, respectively. }
\end{center}
\end{table}

The terms (1-5) can be written as uniaxial anisotropies along the $z$ direction, and are thus redundant. In our model, we retain (1) and (6) to capture the in-plane symmetry breaking, but for simplicity neglect the terms (7-9). Further neutron diffraction experiments are required to clarify which of the planar terms are dominant.

\section*{Appendix B: Monte Carlo techniques}

The helicity modulus, specific heat, and magnetic susceptibility in this work are sampled from corresponding Boltzmann distributions using spin-cluster Monte Carlo. For $J' = 0$, we simulate an $L \times L \times 1$ lattice. For $J' \neq 0$, we simulate an $L \times L \times 4$ lattice. The lattice is subjected to periodic boundary conditions in all directions. We employ Wolff cluster updates \cite{wolff1989} with simulated annealing \cite{kirkpatrick1983}. The effect of the anisotropy field is accounted for by including a ghost spin, following the algorithm presented by Ref~\cite{kentdobias2018}. We generate spin reflections randomly by sampling a Gaussian random vector $\boldsymbol \Gamma$ where $\Gamma_i \sim N(0, 1)$, where $N(\mu, \sigma)$ is a normal distribution of mean $\mu$ and standard deviation $\sigma$. Then, we use the reflection matrix $R = \mathbb{I}_{3 \times 3} - 2 \boldsymbol \Gamma \boldsymbol\Gamma^T / |\boldsymbol \Gamma|^2$, where $\mathbb{I}_{3 \times 3}$ is the $3 \times 3$ identity matrix. 

To measure $\langle A \rangle = \overline{A}(T)$,

\begin{enumerate}
    \item Initialize each spin randomly on the unit sphere. 
    \item Perform $n_{\text{max}}$ = $1.5 \times 10^{6}$ ($5 \times 10^{5}$) cluster updates for $L = 8, 16, 32, 64, 128, 256)$, where for the $n$th update, the temperature is set according to the annealing schedule 
    \begin{align}
        T_n = T_0 \left(\frac{T_{n_\text{max}}}{T_0}\right)^{n/n_\text{max}},
    \end{align}
    where ${T_{n_\text{max}} = T}$ and ${T_0 = 2 J}$.
    \item Perform $10^4$ cluster updates at $T$, sampling $A$ after every update. Label these samples $A_j(T)$. 
    \item Record
    \begin{align}
        \overline{A}(T) &= \frac{1}{L}\sum\limits_j A_j(T), \\
        \overline{A^2}(T) &= \frac{1}{L}\sum\limits_j A_j(T)^2, \\
        \sigma_A(T) &= \bar{A}(T)^2 - \bar{A^2}(T).
    \end{align}
\end{enumerate}

\bibliography{bibliography}

\end{document}